# Tittle: Microbial Engineering to Mitigate Methane Emissions in Ruminant Livestock - A Review

Rehema Iddi Mrutu[1], Kabir Mustapha Umar[1*], Adnan Abdulhamid[2], Morris Agaba[3]

and Abdussamad Muhammad Abdussamad[4]

[1]Center for Dryland Agriculture, Bayero University Kano, Nigeria. P.M.B. 3011, BUK, Kano.

[2]Department of Geography, Bayero University Kano, Nigeria. P.M.B. 3011, BUK, Kano.

[3]School of Life Science and Bioengineering, Nelson Mandela African Institution of Science and Technology, Arusha, Tanzania. P.O.BOX 447 Arusha.

[4]Faculty of Veterinary medicine, Bayero University Kano, Nigeria

**\* Correspondence:**

Corresponding Author Kabir Mustapha Umar

kmumar.cda@buk.edu.ng

P.M.B. 3011, BUK, KANO

**Abstract**

Converting methanogens into acetogens can reduce methane emission in ruminant livestock and promote sustainable animal food production. Ruminants, through enteric fermentation convert fibrous biomass to high quality protein that sustain livelihood globally. In this process however, methanogens produce methane, a potent greenhouse gas, from the by-products of enteric fermentation; carbon dioxide and hydrogen.

The objective of this paper is to provide a comprehensive review of the most recent and promising strategies for mitigating methane emissions in ruminants. Specifically, highlighting the potential of reductive acetogenesis as a viable alternative to methanogenesis. The emergence of Clustered Regularly Interspaced Short Palindromic Repeats (CRISPR) technology, and its exceptional precision in genome editing, further enhances the prospects of exploring this avenue.

Indeed, research in ruminant methane mitigation has been extensive, and over the years has resulted in the development of a wide variety of mitigation strategies ranging from cutting our meat consumption, reducing methane intensity to cutting net emissions. There is no doubt that the concepts of meat alternatives like lab-meat, microbial proteins and plant proteins may produce equivalent emissions. Reducing methane intensity through breeding and diet has been limited by our inability to phenotype ruminants in a high-throughput manner and the intensification of feed-food competition. Although chemical inhibitors have demonstrated effectiveness in manipulating the rumen microbiota to reduce net emissions, their success is constrained in terms of duration and feasibility in grazing system. Progress in making acetogenesis the dominant hydrogen sink in the rumen has been hampered by the thermodynamic cost of the reaction and the limited hydrogen levels in the rumen environment. We believe that CRISPR may allow the dominance of acetogenesis by converting methanogens into acetogens. We propose *Methanobrevibacter ruminantium* to be targeted because it is the dominant methane producer in the rumen.

## 1.0: Methane in Global warming and Climate Change.

### 1.1 Why is methane important?

Methane ($CH_4$) is second to carbon dioxide ($CO_2$) in driving climate change, but dominate how quickly the climate warms due to its greater heat-trapping ability; $CH_4$ warms the earth 86 times more effectively than $CO_2$ (Cain et al. 2019). Unlike $CO_2$ which stays in the atmosphere for hundreds of years, methane has a short atmospheric lifetime of 10 - 12 years (Stavert et al. 2022). Therefore, targeting the emission of $CH_4$ will slow global warming more quickly compared to targeting $CO_2$.

The effect of $CH_4$ has two components; climate warming and atmospheric pollution. $CH_4$ directly warms the earth by its own radiative forcing (RF) of 0.97 Watts per square meter ($Wm^{-2}$). Oxidation of $CH_4$ in the atmosphere by hydroxyl (OH) radicals or nitrous oxide produces ozone ($O_3$), $CO_2$, and water vapor (**Fig 1**) (Rigby et al. 2017) which add weight to the radiative forcing of methane. For instance, on average, oxidation of one molecule of $CH_4$ in the presence of $N_2O$ yields 2.7 molecules of ozone accounting for a 30% increase green-house effect of methane. In addition, compared to $CO_2$, $CH_4$ has much thinner (unsaturated) infrared radiation line which means increase in $CH_4$ has a much larger additional greenhouse effect than an equal increase in $CO_2$ (Bonev and Alexandrov 1993) . In general, it has been estimated that the $CH_4$ - OH feedbacks increase the temperature impacts of $CH_4$ by around 20% (Collins et al. 2018).

3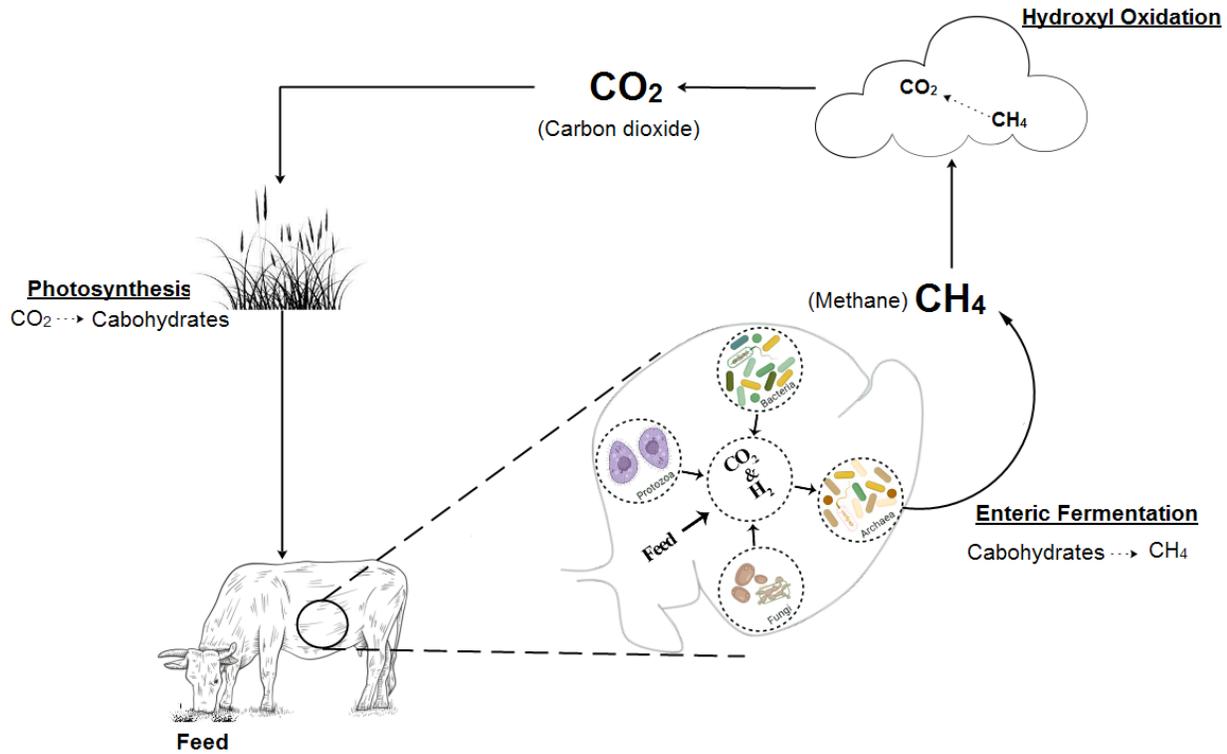

**Figure 1**: Methane in the carbon cycle

The climate pollution effect of methane is mainly caused by ozone; Methane oxidation contributes approximately 15% of tropospheric ozone burden (Collins et al. 2018). Ozone threatens human health, agriculture and ecosystem balance. Globally, about 1 million premature deaths occur annually due to respiratory illnesses resulting from ozone exposure with half of these deaths attributable to anthropogenic $CH_4$ emission. In addition, ozone exposure affects crop growth and productivity. Recently, Mills et al. (2018) reported a loss of 537 million tons of crop yield between the year 2017 and 2019 due to ozone exposure. By putting a dollar value to all these methane damages, it was found that, emitting a ton of $CH_4$ is 50–100% more costly than the corresponding one ton of $CO_2$ emission (Errickson et al. 2021).

The current global emission of methane is approximately 600 tetragram per year ($Tgyr^{-1}$) (Calabrese et al. 2021). This amount is approximately 2.6 times the pre-industrial era emission and account for a 0.5°C of the current global warming. Generally, research suggest that all significant $CH_4$ sources are human driven and largely biogenic in nature i.e., they involve microbial fermentation. Most recently, research has confirmed that emissions from the year 2007 to present are 85% from microbial sources with about half of it coming from the tropics (Basu et al. 2022).

Human derived microbial emissions fall under three areas (**Fig 2**): livestock production (115 Tg $CH_4yr^{-1}$), landfills and waste (68 Tg $CH_4yr^{-1}$) and rice paddies (30 Tg $CH_4yr^{-1}$) (Saunois et al. 2019). Within the livestock sector, emissions from enteric fermentation contribute the largest proportion (85%) amounting to 98Tg $CH_4\,y^{-1}$ (Saunois et al. 2020). Cattle, due to their large population (1.5 billion animals), large rumen size, and particular digestive characteristics, account for the majority of enteric fermentation CH4 emissions from livestock worldwide (Malik et al. 2021). On average a beef cow is estimated to emit up to 500 litres (equivalent to 328.5 grams) of methane per day (**Fig 3**) which exit the animal mainly (95%) through the mouth (Cezimbra et al. 2021) This methane not only does it harm the environment, but also impact animal performance as it accounts for about 12% -15% loss of dietary energy (Tapio et al. 2017).

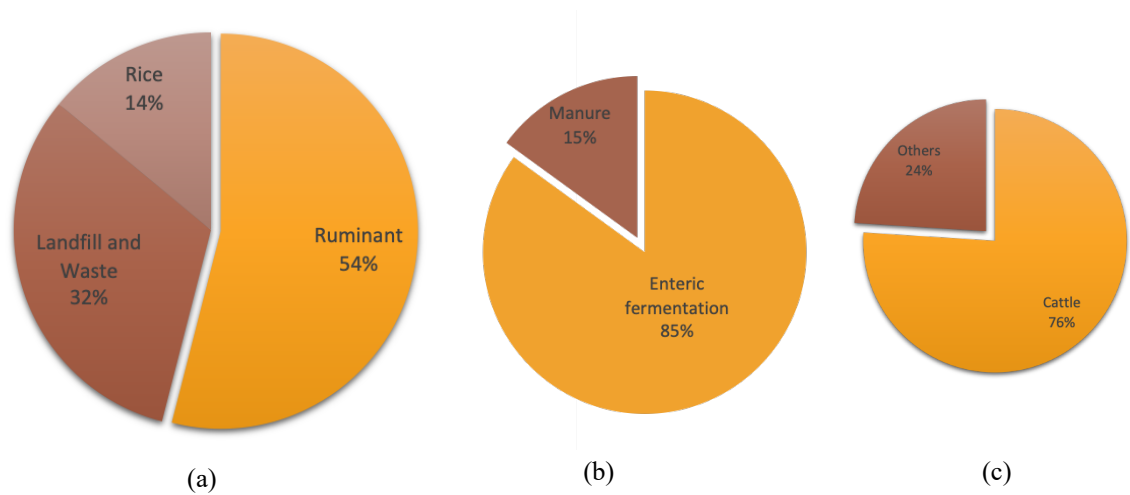

**Figure 2**: Methane microbial emissions by source; (a) human derived microbial sources (b) ruminant livestock emission sources and (c) contribution of cattle in ruminant enteric emissions.

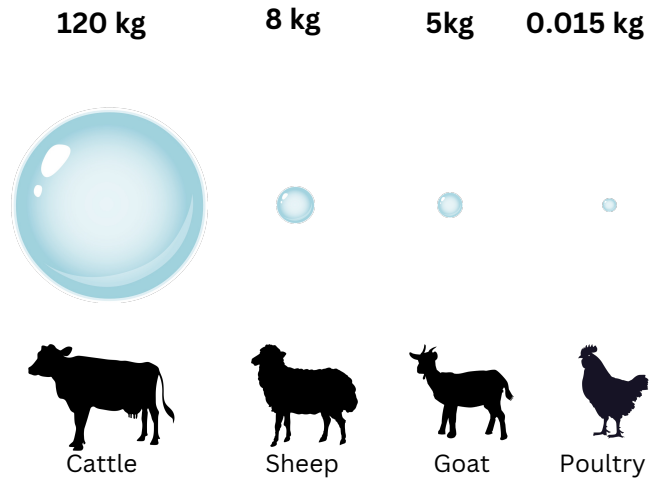

**Figure 3**: Global yearly emissions of methane by different species.

As the population and economic muscles expand, both consumption and wastes will increase microbial emissions. In the case of the enteric fermentation, the continual increase in demand for ruminant products is predicted to increase meat and milk production by 76% and 63% respectively from their levels in 2010 by the year 2050 (FAO, 2018). This extensive production coupled with already existing extreme weather events will exacerbate microbial emissions to levels beyond planetary balance. For instance, there is strong evidence that prolonged dry season in the drylands reduce feed quality and cause feed scarcity both of which reduce digestibility and increase emissions from enteric fermentation (Tadesse and Dereje 2018). On the other hand, prolonged rain seasons expand ruminant feed sources, which in turn promote overfeeding consequently increasing emissions (Schaefer et al. 2016). In landfills and wetlands, temperature rise and rainfall creates and expand ideal conditions for methane-emitting archaea respectively (Moomaw et al. 2018).

Taking advantage of its potency, short life time, the described climatic impact above, and that it is largely microbial driven; methane is considered to be an attractive and cost-effective target for climate mitigation strategies. Targeting methane will give greater mitigation impact because of its GWP and faster mitigation impact because of its short atmospheric stay. If all the livestock enteric methane emissions were stopped, the climate would cool by 0.3 °C by 2045 (Reisinger et al. 2021).

## 2.0 Broad strategies for reducing CH4 emission from ruminants.

In recognition of the dangers and opportunities of methane mentioned above (including being an intense warmer and significantly more reversible), the United Nations Climate Change 26th Conference of Parties (COP26) through the Global Methane Pledge set a goal to decrease agriculture methane emissions by 30% by the year 2030 (Meinshausen et al. 2022). Recently, COP27 launched an initiative called "fast mitigation sprint" the goal of which is to focus on big and correctable sources of methane like livestock sources and to act as quickly as possible (BBC, November 20, 2022).

To deliver against these targets, the two most influential institutions examining emission from livestock, the United Nations Food and Agriculture Organization and the Intergovernmental Panel on Climate Change (IPCC) have recommended a reduction of global cattle herd as a mitigation measure to reduce enteric fermentation and comply with the 1.5°C warming limit of the Paris Agreement ( FAO 2022; IPCC 2022). Several much-cited reports, including the Lancet report (Willett et al. 2019), the Greenpeace report (Greenpeace 2020),  have joined this call by suggesting a 50% reduction in red meat consumption by 2050.

Reduction in meat consumption however, is not practical in numerous ways. First, it conflicts with the World Health Organization (WHO) recommendations to meet some key United Nations Sustainable Development Goals (UNSDGs): e.g., SDG2 on nutritional requirements and food security and SDG1 on poverty reduction. This is especially relevant in low-income countries where animal protein sources are still insufficient (Asano and Biermann 2019) and their consumption expected to rise as population and average individual income increases. Reducing animal numbers in these societies translates to the challenge of meeting their nutritional demand mainly, vitamin B12, vitamin D, riboflavin, iron, calcium, and zinc (Fehér et al. 2020) which will trigger health problems such as malnutrition, stunted growth or anemia. Secondly, these communities' livelihoods and economies are reliant on cattle production, meaning banning cattle production will affect their general income, asset provision and insurance (Humpenöder et al. 2022). Thirdly, if cattle production is reduced at a large scale, this would also mean that alternatives for fertilizer (that comes from feces, bones and blood), leather products (originating from hides and skin) and pet foods (from organs) must be put in place.  Replacing these with synthetic products would add more GHGs to the environment especially through production, processing and transportation (Cheng, McCarl, and Fei 2022).

 Again, it is generally assumed that the land used for livestock production could alternatively be used as carbon sink through forestation. However, where livestock farming is largely practiced (for instance in dryland savannas, parklands, deserts, steppes, Arctic tundra, Mediterranean hills and plains or mountains) land use alternatives like crop farming and tree growing are usually not feasible (Houzer and Scoones 2021). This means that the areas will turn into bushes that attracts bushfires further exacerbating GHGs emissions. Pastoralism alone covers over half of the world's land surface (Scholes 2020). If this land is vacant of animals, it might produce equivalent emissions, if not more. Livestock on the other hand, have a better capacity to store carbon in the ecosystem by suppressing bushfires and locking carbon into a much safer place- the soil.

## 2.1 Suggested alternatives to food animal proteins

Shifting human diets to plant-based foods will obviously moderate the demand for animal food products consequently cutting down methane emissions. In terms of production, plant-based diets have lower environmental foot print than animal foods (Springmann et al. 2018). For instance, to make a kilo of protein from kidney beans requires approximately eighteen times less land, ten times less water, nine times less fuel and twelve times less fertilizer in comparison to producing 1 kg of protein from beef cattle (Sabaté et al. 2014). Studies have shown that a 70% reduction of GHG emissions and land use could be achieved by shifting our diets to plant-based proteins (Fehér et al. 2020).

Whilst plant-based production is considered safe, their consumption may lead to more emissions and are not feasible economically and health wise. Similar to the rumens of animals, humans produce methane from metabolism of methanogenic species particularly hydrogenotrophic methanogens (Polag and Keppler 2019b). Up to 15 species of methanogens have been identified in human gastro intestinal tract (GIT) and contribute to the production, on average, of 0.35 litres (equivalent to 0.23 grams) of methane per person per day (Polag and Keppler 2019a; Djemai, Drancourt, and Tidjani Alou 2021). In 1986, when the human population was 4.7 billion, total human methane production via breath only was estimated to be 0.3 Tg per year (Crutzen, Aselmann, and Seiler 1986). Assuming a direct correlation to population, by now human emit 0.51 Tg per year considering 8 billion people in 2023. For comparison, this amount is significantly lower than emissions from ruminant animals. However, it must be noted that these human estimations represent only a fraction (25%) of what humans emit, leaving behind the 75% that goes through flatus which is usually difficult to measure and under reported. Further they do not consider various factors that influence human emission like age, sex, food and geographical location. With this in mind; If a transition to plant-based diet is made these numbers will increase substantially because digestion of fiber rich diets produces lots of methane than fermenting protein or fat rich diets (Wilson et al. 2020). The very same concept that explains why herbivores emit more methane than omnivores. Therefore, it will be a matter of replacing 1.5 billion cattle with 8 billion emitting humans.

In an effort to determine the effect of plant-based diet on methane emissions, a recent study looked at the correlation of methanogen abundance with diet in humans. Based on their analysis Wilson et al. (2020) and colleagues found evidence that western diet which is largely dominated by meat, fish, eggs and milk products, negatively correlates with the abundance of methanogenic community. This confirmed data of Borrel et al. (2020), that rural Africa produces high amount of methane (above the 1- parts per million (ppm) threshold) than western adult population. Indeed, majority of households in rural Africa are vegetarians by force (economic, traditional, religious and cultural forces) as evidenced by household and demographic surveys conducted in Sub Sahara Africa between 2018, and 2021 showing that 90% of diets in the rural are plant based (Kyomuhendo and Adeola 2021; Mensah et al. 2021). With these findings, a switch to plant-based diet might not guarantee a significant reduction in emissions.

Moving forward, land related GHG emission savings from plant-based diet production need to be contrasted with emissions cost of consuming plants before switching to vegetarianism. Countries that have achieved carbon neutrality status like Gabon and Bhutan and countries where vegetarianism is highly practiced like India, Italy, UK, Germany and Netherlands (Fehér et al. 2020) could also serve as a natural experiment to prove if vegetarianism works.

### 2.1.1 Cultured meat

One of the most promising alternatives to animal source protein is cultured meat or the so-called lab meat. Cultured meat bypasses the extravagant length of rearing a whole animal and is currently available in the market (Ismail, Hwang, and Joo 2020). Here a steak is produced from just one fat or muscle cell using stem cell technique and tissue engineering the details of which are described extensively elsewhere (Stephens et al. 2018; Stephens and Ellis 2020). By just removing the animal from the equation, GHG emissions, land and water use are expected to reduce substantially. However, converting a cell into a steak in the lab costs a fortune. For instance, in 2013, researchers from Maastricht University spent $2,470,000/kg of production just for a proof-of-concept (Rubio, Xiang, and Kaplan 2020). The question then is how much would it cost if the attempt is done to scale up the production? Furthermore, research has constantly reported the difficulty in balancing mechanical properties and nutritional parameters (Rubio, Xiang, and Kaplan 2020). Nutrients like vitamin B12 and flavors are constantly supplemented to achieve whole animal nutrient profiles.

### 2.1.2 Microbial protein (MP)

Currently the only commercially available protein sources that are classified as "complete" come from fungus. These fungi, primarily *Fusarium venenatum*, are cultivated in heated bioreactors using sugar as feedstock (Humpenöder et al. 2022). The microbial biomass has high protein content equivalent to ruminant meat and safe for human consumption. Moreover, life cycle assessment (LCA) studies argue that for each unit of ruminant meat substituted with mycoprotein there will be a reduction of GHG emissions by 80%, water use by 90% and land use by 90% (Humpenöder et al. 2022). However, this assessment does not take into account the energy in form of electricity used to run the bioreactors. Therefore, if we contrast the land-related GHG emission savings of MP with the energy-related GHG emissions we find that mycoprotein production costs more energy than conventional ruminant meat production. In fact, one study reported that energy demand for MP is 0.03–25 times that of other feed protein (Järviö et al. 2021). Unless renewable energies like solar and wind become highly efficient and cost effective, and eventually replace the existing energy system, MP will continue to have huge environmental impact. To summarize, taking into account the economic feasibility of these technologies, their energy requirements and consumers acceptance, it may be more feasible to transform meat production rather than replace it entirely. If animal rearing could be integrated with all these techniques, sustainability can be achieved.

Considering the challenges involved in achieving significant reduction in livestock population and consumption of animal products to mitigate emissions, it is necessary to develop alternative strategies that can meet emission reduction targets without compromising traditional livestock practices. Proposed approaches can broadly be categorized into two groups: (1) strategies to reduce emission intensity per kilogram of animal produce and (2) strategies to reduce net emissions by targeting microbial fermentation.

## 2.2 Reducing emission intensity per kg of animal protein produced

### 2.2.1 Management (husbandry) Practices

To sustainably meet the ever-growing demand for animal foods, reducing emission per unit product is critical (Richardson et al., 2021). Meaning that less inputs need to get to the system to make a kilo of meat or milk. Management practices i.e., genetic selection, advances in nutrition and disease control combined with advanced animal reproduction technologies i.e., artificial insemination and embryo transfer have resulted in a 30% increase in global livestock productivity. Nutritional advancement has accrued mainly through the use of grain and feed supplements. Due to higher nutrient density and easy digestibility, grains fasten animal growth rate and reduce animal residence time which means animals emit less emissions over the course of their lives. For instance, in the United States, the average annual milk yield in grain-finished cow moved from 1,890 kg to 9,682 kg of milk per cow between 1924 and 2011 and the carbon foot print reduced by 41% between 1944 and 2007 (Capper and Bauman 2013; Georges, Charlier, and Hayes 2019). Nutritional management is especially attractive in developing world where the room for improvement is greatest as these countries develop their animal resources to meet increasing demand. However, a shift to western management system will introduce competition with human food supply as more land would be needed for grain production to accommodate both human and animal food which could create additional emissions. For this reason, research must focus in identifying nutrients with the potential to reduce emissions without compromising human food systems.

### 2.2.2 Breeding low emitting cattle

There is a natural variation in the amount of methane produced per kilogram of dry matter consumed across animals. This trait is moderately (0.13-0.35) heritable in cattle and sheep which makes breeding for low emitting animals a possibility (Black, Davison, and Box 2021). To breed for low emitting animals, a stock population that will propagate the next generation must be carefully selected. To aid in the selection of seed stock population massive phenotypic and genotypic analysis of the methane trait is required. The major difficulty of using phenotypic records for selection is to distinguish between the impact of genes from that of the environment (Georges, Charlier, and Hayes 2019). Furthermore, the feasibility of measuring methane emission amount and patterns appears to be low because of expensive equipment and intensive animal handling (González-Recio et al. 2020). Identification of methane emission proxies could help offset the cost of methane measurement and accurately select for low emitting animals (Lassen and Difford 2020). The diary sector too, has the potential to offset the cost of measurement because advanced animal reproduction technologies like superovulation and embryo transfer allows measurement on just a subset of progenies which technically reduce the numbers of animals to be measured (González-Recio et al. 2020).

Gene based selection on the other hand is an alternative to the time consuming and costly phenotypic measurement. It can improve the effectiveness of selection even at low heritability levels. Again, the diary sector represents a unique opportunity to implement this approach because of well-established genotypic datasets. Generally, the potential for genetic improvement through direct selection for methane trait is limited at 0.2 – 0.4% per year, which collectively has the potential to reduce only 4-8% of methane over 20 years (Black, Davison, and Box 2021). This means that lowering global emission by ruminants is unlikely to be realized within the timeframe of action needed to keep global warming below 1.5°C. The dairy industry however has the potential to improve the rate of genetic gain for methane mitigation because of the wide use of artificial insemination which allows rapid gene transfer.

Overall, breeding interventions are slow. They take generations for a global impact to be realized. Furthermore, they compromise biodiversity and conflicts with other functional traits (Manzanilla-Pech et al. 2021). González-Recio et al. (2020) showed that higher emitters tend to yield larger amounts of milk. This aligns with the comment made by de Haas et al. (2017) that it is possible to get an improvement for environmental impact of 20% to 34% at the cost of a decline in milk production of 7% and in functional traits of 5%. This observation suggests that breeding for low emitters will compromise animal productivity. Moving forward, research should focus on controlling the environment because managing the environmental factors have shown bigger effect on emission which makes the challenge more tractable in a shorter period than gene-based animal breeding

## 2.2.3 Capture of methane produced by animals and their waste

Whilst there has been significant advancement in capturing methane from sources such as landfills and rice paddies, technologies to capture methane from ruminant livestock are poorly developed. The energy content in methane is estimated to be 50.4 Megajoules per kilogram (MJ/kg) (Angelidaki et al. 2018). In principle, just 12 cows are sufficient to provide an average household with its daily domestic gas as per Crutzen, Aselmann, and Seiler (1986). In practice, however, it is usually difficult to capture ruminant methane because each animal exists as an individual source that roam around in the field. Technological solutions to capture most (95%) of burped methane are not only fully developed but also interferes with animal welfare (Kaya and Kaya 2021). These observations have pushed interest in developing tools to capture the 5% of methane that exit with manure. Technologies like biogas digesters are commonly used in developing world at the level of households or small farming communities to capture this methane. Basically, a mechanism is set to collect manure which is then allowed to ferment, the gas is harvested for energy purposes while the manure is used on farms (Sarah, Susilawati, and Pramono 2021). This technology has improved the livelihoods of rural communities by providing a reliable source of renewable energy and improved soil fertility. The efficiency of this system however, is hampered by the mix of methane with other gases like $CO_2$ and poor fermentation techniques (Angelidaki et al. 2018). To enhance the efficiency of bioreactors, $CO_2$ from the mix of the biogas must be removed. More recently the use of hydrogenotrophic methanogens and algae which convert the $CO_2$ in the mix into more methane has been tested and proved to be effective (Nagarajan, Lee, and Chang 2019). The biggest challenge with these technologies is the availability of cheap $H_2$ sources and establishment of a cost-effective microalgal cultivation system respectively. Therefore, the concept of water electrolysis using renewable electricity to split water into hydrogen and oxygen is an attractive area of future research. In light of the challenges above, the solution to ruminant methane lies in reducing enteric methane generation in the rumen.

## 3.0 Rumen microbial dynamics

Livestock, particularly ruminants, use low quality fibrous feedstock and convert it into high quality livestock products including meat and milk. They are able to achieve this via the action of a symbiotic microbial ecosystem in their GIT (**Fig 1**) (Martínez-Álvaro et al. 2022). Essentially, microbial fermentation of the feedstock in the rumen produces short chain fatty acids (SCFAs) which are consumed by the host for its own energy and growth. Some of the rumen microbiota pass down the GIT where they are themselves digested to further nourish the host with protein, long chain fatty acids, and vitamins (Mizrahi, Wallace, and Moraïs 2021). Concomitant with the nourishing effects, the fermentative action of rumen microbiota produces greenhouse gases, most notably methane.

The rumen microbiota is composed of archaea, bacteria, fungi, and protozoa (**Fig 1**) (John Wallace et al. 2019). Bacteria are the most numerous and diverse comprising 95% of the rumen microbiota (Pereira, de Lurdes Nunes Enes Dapkevicius, and Borba 2022). Protozoa are the most abundant by biomass (Solomon et al. 2021) and their diversity and abundance tend to fluctuate more widely across breeds and feed type (Newbold et al. 2015). Numerically, fungi comprise a very small component of the rumen microbiota but they are extremely efficient at degrading the toughest plant material because they house an extensive set of enzymes and rhizoids which penetrate plant structural barriers subsequently increasing the plant cell surface area for other microbes to colonize (Xue et al. 2020). Although archaea are the least abundant (Friedman et al., 2017), they produce all the methane (**Fig 1**) (John Wallace et al. 2019). Bacteria and archaea show significant heritability and form the key core rumen microbiota (Martínez-Álvaro et al. 2022) that is critical for rumen viability and animal development.

The rumen microbiota is a dynamic ecosystem with intricate interdependences. Some bacteria, fungi and protozoa digest plant fiber and extract energy by anaerobic respiration; the products of primary fermentation are passed down the food chain and are ultimately degraded to hydrogen and carbon-dioxide (Lyu et al. 2018). Although $H_2$ and $CO_2$ can be expelled from the rumen by eructation (burping), the energy in the mix would be lost to the rumen ecosystem. Consequently, some archaea and some bacteria are able to use the $H_2$ and $CO_2$ for energy conservation and growth and in turn produce methane and (or) acetate respectively (Ma et al. 2021).

This complex interdependence between the key hydrogen producers and key hydrogen utilizers has been repeatedly observed with methanogens. For instance, between the rumen fungus *Neocallimastix frontalis* and *Methanobacterium formicicum* (Nakashimada et al. 2000); between the bacterium *Ruminococcus* and *Methanobrevibacter ruminantium* (Henderson et al. 2015); and between protozoans *Entodinium*, and *Methanobrevibacter thaueri* (Xia et al. 2014). The abundance of these hydrogenic organisms has been linked to methane emission and animal production; depleting them from the rumen reduces methane emission (Ibrahim et al. 2021) but reduces animal performance. In this context, research efforts should focus on identifying potential hydrogen utilizers that can replace methanogens and be used as rumen modifiers for methane mitigation.

Generally, the dynamic of rumen microbiota is largely influenced by three factors; feed (Wilkinson et al. 2020), age of the animal (Martínez-Álvaro et al. 2022) and genetics (Yáñez-Ruiz, Abecia, and Newbold 2015). Feed type and amount has the most dramatic effect on microbial dynamics and the subsequent effect on methane generation. Poor quality feed with high fiber content have hugely been associated with high methane emissions (Vaghar Seyedin et al. 2022) than highly digestible low fiber feed (Hayek and Garrett 2018). Ease to digest feed, increases passage rates which in turn wipe up methanogenic population resulting in total methane reduction. Additionally, because fermentation happens so fast, hydrogen build at a higher pace resulting in higher rumen hydrogen concentration. Naturally when hydrogen levels are too high the system shifts carbohydrate fermentation towards hydrogen consuming reaction like propionate, to keep things in balance (Arndt et al. 2023).

Bacteria like *Firmicutes* and *Ruminococcaceae* have been found to dominate rumens of high emitters (John Wallace et al. 2019; Furman et al. 2020). Since these are key fiber degraders and highly associated with hydrogen production in the rumen, it is reasonable to assume high fiber intake in these animals. On the hand, considering that *Bacteroidaceae*, *Succinivibrionaceae*, *Quinella* spp and *Dialister* are commonly found in the rumen of cows that emit extremely low levels of methane (John Wallace et al. 2019; O'Hara et al. 2020), it can also be concluded that these animals have less fiber intake because these microorganisms are efficient in utilizing hydrogen resulting into propionate production. Bacteria that are capable of producing lactate and succinate, such as *Fibrobacter spp.*, *Kandleria vitulina*, *Olsenella spp.*, *Prevotella bryantii*, and *Sharpea azabuensis* also dominate the rumens of low emitting animals (Wallace et al. 2015). An intriguing additional observation is seen with Tammar wallaby, a herbivores foregut fermenter similar to ruminant, which exhibits remarkable abundance of *Succinivibrionaceae* community. This microbial family is believed to play a role in the wallaby's significantly lower methane production, which is only around one-fifth that of ruminant per unit of feed intake (Pope et al. 2011). These bacterial markers could be predictive of high and low methane phenotypes in ruminant animals and therefore more studies are needed to strengthen this relationship.

The microbial composition and the subsequent effect on methanogenesis also change with age. Young animals have less diverse, less abundant and quite fluid microbiota (Cammack et al. 2018). This together with their small rumen size and their diet accounts for their small methane emissions. The fluidly nature of young animals' microbiota, potentially presents opportunities for manipulation of the ecosystem for reduced methane emissions. As the animal matures, the microbiota stabilizes and becomes more diverse and abundant to support methanogenesis. Cattle achieves this stabilization three weeks after birth (Yáñez-Ruiz, Abecia, and Newbold 2015). Research have thus suggested that, beneficial microbial signatures must be imprinted within this sensitive (first three-weeks) window for the animal to have a long term carry over effect (O'Hara et al. 2020).

Besides feed and age, host genetics is also a main determinant of microbial composition and methane emissions. Both animal species and breed influence microbiota composition and consequently methane production (Roehe et al. 2016). One of the current exploratory efforts to understand animal genetics involvement in microbial composition discovered a significant association between a region on chromosome 6 with the densities of *Euryarchaeota*, *Actinobacteria* and *Fibrobacteres* (Golder et al. 2018). Despite this achievement more studies are needed to not only establish the association but also determine the extent of the influence and the underlying mechanisms. Overall, rumen microbial composition and the factors that influence their abundance, diversity and function present opportunities for microbial manipulations and can be used as potential tool in methane mitigation strategy.

## 3.1 Targeting rumen microbiota

### 3.1.1 Suppression of methanogenesis using chemical Inhibitors

Targeting methanogens and the subsequent effect of methanogenesis is the most viable approach towards net methane reduction in ruminant animals owning to its positive effect on fermentation, metabolism and production. If methanogenesis is inhibited the 12% energy loss, can be captured and channeled towards host and microbiota nutrition subsequently increasing animal performance (Karekar and Stefanini 2022).

A recent meta-analysis, which examined the impact of all chemical inhibitors in an *in vitro* experiment, identified macroalgae and 3- nitro oxypropanol (3- NOP) as the most potent suppressors of methanogenesis (Kebreab et al. 2023). The use of macroalgae to reduce enteric methane emission in ruminant livestock has seen a significant surge in interest in recent years (Karekar and Stefanini 2022; Ábrego-gacía et al. 2021). Several reports worldwide have investigated the potential of different kinds of macroalgae including red, brown and green species. Although all of them have shown promising results, *Asparagopsis spp* (*A. taxiformis* and *A. armata*) have shown significant mitigating effect in *in vitro* experiments with rumen fluid (Kinley et al. 2016).

Research suggests that, the anti-methanogenic effect of *Asparagopsis spp* is derived from its low molecular weight halogenated compound, with the brominated halomethane known as bromoform being the most prevalent (Roque et al. 2021). However, bromoform has been reported to be carcinogenic and contributes to ozone layer depletion (A. Patra et al. 2017). Despite this, bromoform has demonstrated the ability to significantly reduce methane emission by up to 80%. If research addresses the carcinogenic effect of bromoform, macroalgae have a great chance of adoption by farmers as it increases animal performance; about 14% feed conversion efficiency was observed in beef cow following macroalgae supplementation (Roque et al. 2021).

Contrast to carcinogenic effect of macroalgae, 3- NOP has been consistently reported to be safe for both animals and consumers (Y. Zhang et al. 2021). 3- NOP which is currently in the markets, is an analogue of the enzyme methyl- coenzyme M reductase (*Mcr*) which facilitate the final step of methane formation. Up to 30% reduction in methane emission in total mixed ratio farm system have been observed without any toxicity (Wesemael et al. 2019; X. M. Zhang et al. 2020). Despite this, the overall improvement in animal performance is limited presumably due to accumulation of hydrogen gas that would otherwise be used in generation of methane. Nonetheless, recent research suggests that animal productivity, can be increased by supplementing 3 NOP with phloroglucinol, which captures excess hydrogen to produce beneficial metabolites for the host (Dijkstra et al. 2018). Moreover, since acetogenesis is an alternative hydrogen sink in the rumen, 3 NOP can be used in combination with acetogens to capture the hydrogen and convert it into acetate.

Overall, the effectiveness of chemical inhibitors in the rumen is limited because of their short stay. To address this issue, ongoing research aim to develop slow-release formulations and apply them to young animals to promote long term effects (McCauley et al. 2020). However, this technology may not be applied in developing countries, where improving animal productivity remains a major challenge.

### 3.1.2 Biological control of methanogens by means of phages

Rumen phages play a key role in regulating the populations of microbiota in the rumen. They infect and replicate within specific species, causing their lysis and reducing their numbers, consequently creating a well-balanced microbial community important for proper digestion and overall animal health (Lobo and Faciola 2021). For this reason, phages can be used as a tool to manipulate the rumen microbial community in a targeted manner. This has stimulated the isolation and characterization of rumen phages for biocontrol purposes. Whilst the first isolation was done as far back as 1966, to date, only few rumen phages have been well characterized (Leahy et al. 2010). This is due to the inherent difficulties in culturing rumen microorganisms. Of the phages that have been successfully isolated, majority belong to the families *Myoviridae*, *Siphoviridae*, *Mimiviridae*, and *Podoviridae* (Lobo and Faciola 2021) which are mostly bacterial phages and highly associated with dominant rumen bacterial phyla like *Firmicutes* and *Proteobacteria* (Namonyo et al. 2018). Recently, five bacterial phages have been isolated from rumen samples that target the predominant fibrolytic ruminal bacterium *Butyrivibrio fibrisolvens*. Other studies have also investigated phages such as phages φBrb01 and φBrb02, φSb01, φRa02 and φRa04 that specifically target *Bacteroides, Streptococcus* and *Ruminococcus,* respectively (Gilbert et al. 2017). By targeting bacteria, substrate hydrogen for methanogenesis becomes limited subsequently suppressing methanogenesis.

Few phages that directly target methanogens have been studied as well. This includes those that infect *Methanothermobacter wolfeii* (Schofield et al. 2015) *Methanothermobacter marburgensis* (Liesegang et al. 2010) and *Methanosarcina mazei* (Weidenbach et al. 2017). Additionally, a prophage Q-mru and its lytic enzyme endoisopeptidase PeiR were discovered through genome sequencing of *Methanobrevibacter ruminantium*, a common rumen methanogen. The lytic activity of PeiR have been tested on pure cultures and results confirmed a dramatic (97%) drop in culture density and a reduced methane yield.

Although lytic phages can be effective in targeting specific organisms in the rumen, their specificity can also limit their effectiveness as non-targeted organisms can recolonize and repopulate the rumen environment (Morkhade SJ, Bansod AP 2020). This challenge highlights the need for mass identification of rumen phages to facilitate the development of broad-spectrum phage treatments. However, progress in this area has been hampered by the inability to culture rumen microorganisms. Fortunately, with the decreasing cost of sequencing, multiple genomes of archaea can be sequenced to help identify ruminal viruses and facilitate the development of broad-spectrum phage treatment. In addition, diet regimen can also be used to enhance virus population and manipulate microbial lysis in the rumen, as reported by (Anderson, Sullivan, and Fernando 2017).

### 3.1.3 Promotion of alternative consumption of $H_2$ and $Co_2$

Since methanogenesis is influenced by intracellular and intercellular flows of metabolic hydrogen (Ungerfeld 2020), targeting both its concentration and direction of flow will indirectly influence methanogenesis. Methods that decrease hydrogen production by lowering the number of hydrogen-producing gram-positive bacteria, fungi and protozoa in the rumen have been tested (Kelly et al. 2022). Fatty acids for instance, blocks rumen fermentation subsequently suppressing the abundance of fermentative community of bacteria like *Ruminococcus albus*, *R. flavefaciens*, *Neocalimastrix spp.*, and *Desulfovibrio* (Min et al. 2022), protozoa like *Entodinium caudatum* (Darabighane et al. 2021) and fungi in favor of propionate production. This together with the fact that fats are not fermented in the rumen produce less substrate hydrogen for methanogenesis (Belanche et al. 2020). Methane-suppressing effect of fatty acids however is not consistent and largely dependent on the type of fat, concentration and nutrient composition of the diet fed to the animal. In cattle, a 15% reduction in methane was observed (**Table 2**) when the diet contained 6% polyunsaturated fatty acids and low fiber. But then 6% of 15 kg of hay (the usual dry matter intake of a cow) equates to almost a kg of oil (0.9 kg) every day to a cow that weighs 500 kg. As the globe suffers from vegetable oil crisis partly due to the Russia Ukraine war, more sustainable approaches that do not introduce competition with human food supply are critically needed. Moreover, the massive production of oilseed comes at a high cost in terms of GHG emissions and biodiversity destruction. Compared to concentrate feeds production, oilseed production emit nearly as much upstream GHG emissions per kilogram of dry matter (1.27 vs 0.70 $CO_2$ equivalents kg dry matter $^{-1}$) (Arndt et al. 2023).

| Strategy/Approach | Methods | Mode of action | Success rate | Reference | Limitation | Gaps and Future research |
|---|---|---|---|---|---|---|
| Demand management | Plant based diet<br>Meat substitute | Reduce cattle heard | Undetermined | IPCC (2018:327)<br>(FAO, 2018) | Behavioural change<br>Conflicts SDG2, and SDG1 | LCA studies to contrast emission savings from plant-based diet production to emission cost of consumption<br>Quantify human emissions based on diet |
| Reduce methane intensity | Breeding for low emitters<br>Advanced husbandry management | Improved animal productivity | 8% | (Black, Davison, and Box 2021) | Slow<br>Cost of measuring methane<br>Compromise diversity<br>Relationship between traits | Develop cost effective methane measurement tools<br>Studies to identify proxies<br>Studies to establish relationship between traits |
| Reduce net emissions | Chemical inhibitors:<br>    Macroalgae<br>    3NOP<br><br>Biological inhibitors:<br>    Phage<br><br>Targeting substrate $H_2$<br><br>    Concentrate<br><br>    Fats<br><br>    Acetogens | Inhibit methanogenesis | 80%<br>30%<br><br><br>97%<br><br><br><br>30%<br><br>15%<br><br>5% | (Roque et al. 2021)<br><br>(Duin et al. 2016)<br><br>(Altermann et al. 2018)<br><br>(Ungerfeld 2020)<br><br>(Williams et al. 2020)<br>(Muñoz et al. 2021) | Short term effect<br><br>Non targeted *spps* recolonization<br><br>Rumen Acidosis<br>Feed-food competition<br>Limit digestion<br><br>Limited hydrogen<br>Thermodynamics | Develop slow-release formulations<br>Development of broad-spectrum phage treatments<br><br>Discover means to address acidosis other than antibiotics<br><br>Select and promote high hydrogen producers<br>Identify competitive acetogens<br>Engineering WLP |

**TABLE 1:** Trends in Rumen Methane Abatement and Gaps/Future Research

By contrast, a more useful strategy would be to redirect hydrogen to already existing rumen fermentation pathways like propionate, nutritionally useful to the animal and its microbial consortium. This way hydrogen producers will be maintained and they will shift fermentation away from formate, lactate and ethanol toward acetate, and increased cellulose digestion (Ungerfeld 2020). Propionate is the main glucose precursor in ruminants and therefore important to animals with high requirements for glucose such as high producing dairy cows in early lactation (Ungerfeld 2020). But for propionate pathway to compete with methanogenesis animals must be fed diets rich in starch. Starch is easily fermented in the rumen and therefore increases ruminal passage rate which disfavor methanogens growth subsequently shifting fermentation towards propionate (Tseten et al. 2022). However, starch fermentation has been reported to overwhelm the digestive physiology of cattle leading to a condition known as ruminal acidosis (Elmhadi et al. 2022). In order to avoid the occurrence of lactic acidosis, high-grain diet often included antibiotics such as ionophores, which are effective in inhibiting the growth of Gram-positive bacteria that produce lactate. However, the use of antibiotics and growth promoters in livestock production has been restricted worldwide due to concerns about the development of antibiotic resistance. The European Union for instance has imposed a ban on their use since 2003. Therefore, alternative strategies are needed to modulate rumen microbial fermentation while preventing rumen acidosis.

Unlike propionate pathway, the reduction of nitrate (Feng, Bannink, and Gastelen 2020) and sulphate (Zhao and Zhao 2022) can effectively outcompete methanogenesis in the rumen. Despite several prevalent nitrate and sulphate reducers in the rumen environment (Mizrahi, Wallace, and Moraïs 2021), electron acceptors are naturally limited, which restricts these pathways from taking place (Granja-Salcedo et al. 2019). Presumably because nitrate and sulphate are rich in oxygen which can turn the rumen to aerobic fermentation with overall negative rumen function. Studies have shown major reduction in methane emissions with nitrate supplementation. Despite this, several *in vivo* studies have shown nitrate supplementation reduces methane production by 12% in beef steers (Alemu et al. 2019; Granja-Salcedo et al. 2019; Feng, Bannink, and Gastelen 2020), 17% in dairy cows (Meller et al. 2019), 26% sheep (Villar et al. 2019) and 6% goats (Zhang et al. 2019). However, the released intermediates such as nitrite can be toxic for the animal and a great concern to the environment as it can increase enteric and manure nitrous oxide ($N_2O$), a GHG more potent than methane and $CO_2$ combined.

### 3.1.4 Converting the rumen from biogas (CH4) producer to acetic acid generator.

The rumen has the potential to utilize acetogenesis as an alternative pathway for hydrogen disposal in place of methanogenesis. However, this pathway is not frequently observed due to its high thermodynamics cost. Gibbs free energy for the overall reduction of 2 moles of $CO_2$ to 1 mole of acetate is approximately -95 kilojoule per mole (kJ $mol^{-1}$), whereas -130 kJ $mol^{-1}$ is spent for the same reaction to head towards methanogenesis (Kim et al. 2020). Moreover, the low hydrogen affinity of acetogens necessitates a large amount of hydrogen which is naturally unavailable in the rumen.

Contrast to the rumen, in other anaerobic environments like termites (Yang 2018), kangaroos and wallabies (Gagen et al. 2010), acetogenesis is the dominant hydrogen sink. The dominance of acetogens in these environments is partly explained by the abundance of $H_2$- producing protozoa and spirochetes (Karekar and Stefanini 2022) which doubles every 24–48 h (Breznak, Switzer, and Seitz 1988), way faster than hydrogenotrophic methanogens. (Gagen et al. 2014) isolated acetogen from tammar wallaby, that required $H_2$ concentration above 5 micrometre (μM) for reductive acetogenesis which is significantly higher than the 0.95 μM needed by *Methanobrevibacter smithii*, a dominant methanogen in the rumen (Kral et al. 1998; Rosenberg et al. 2013). These findings provide an explanation for the failure of studies using acetogenic probiotics to increase rumen acetogens. Few *in vivo* studies to date, including studies conducted by Kim et al. (2018, 2020) have managed to increase acetate production following supplementation of acetogenic bacteria. In another example methanogens were firstly suppressed before introducing acetogens as demonstrated in a study conducted by Pereira, de Lurdes Nunes Enes Dapkevicius, and Borba (2022) using *Eubacterium limosum*.

Despite the limited success in these studies, there is potential for acetogens to be utilized as a tool to reduce methane emissions in ruminants. There are three possible strategies to consider; 1) increase hydrogen production within the rumen. This can be achieved by selecting for microorganisms with high hydrogen production under rumen condition. Prior to examining this approach, it is crucial to examine a number of questions (**Table 2**). For instance; what is the relationship between hydrogen bacteria in the rumen and the methanogens or acetogens? Certain studies indicate a mutualistic relationship where there is a close association (physical) between hydrogen generators and the hydrogen users. Now, is it feasible to selectively disrupt the relationship between $H_2$ producers and methanogens in the rumen? Furthermore, what are the primary hydrogen generators in the rumen? And do they have any association with acetogens? If not, can we deliberately initiate a syntrophic relationship between them so that hydrogen is more available for acetogenesis? 2) Selection of acetogens with very high efficiency of acetate production at lower hydrogen concentration than ordinarily seen in the rumen. This may be achieved by genetic engineering to increase efficiency of the WL pathway or by employing traditional mutant selection approaches used in microbiology. 3) Genetic engineering of methanogens to abate the methane production pathway (branch) of the WL pathway.

- At what concentration of hydrogen does acetogenesis become activated and maintained?
- At what hydrogen concentration does methanogenesis become impossible?
- How do methanogens and acetogens compete under various hydrogen substrates regimes?
- Is it possible to boost acetogens hydrogen affinity?
- Is it possible to establish acetogens physical interaction with hydrogen producers? Or detach permanently methanogens from protozoans?
- What will happen if ruminants learn from termites' way of digestion?
- If the metabolic capabilities of acetogens is limited to $H_2$ and $CO_2$, would this promote their competitiveness?
- Under what cultivation condition would methanogens produce acetate as sole product/ would methanogens utilize the Acetyl CoA pathway for energy conservation?

**TABLE 2**. Outstanding Questions and Future Research

*3.1.4.1 Genetic engineering of methanogens to abate the methane production pathway*

A significant (77%) amount of methane produced in the rumen arises from bioconversion of the products of fermentation, hydrogen and carbon dioxide; the so called hydrogenotrophic methanogenesis (Subedi et al. 2022). But there is a natural alternative pathway for bioconversion of $CO_2$ and hydrogen to acetate through the Wood–Ljungdahl (WL) pathway. This pathway is commonly employed both by methanogens and acetogens for carbon fixation and by acetogens also for energy generation (Rosenbaum and Müller 2021). The production of acetate from $CO_2$ and $H_2$ is here termed primary acetogenesis; whereby $CO_2$ molecules are sequentially reduced under two branches i.e., the carbonyl and methyl branch (**Fig 4**).

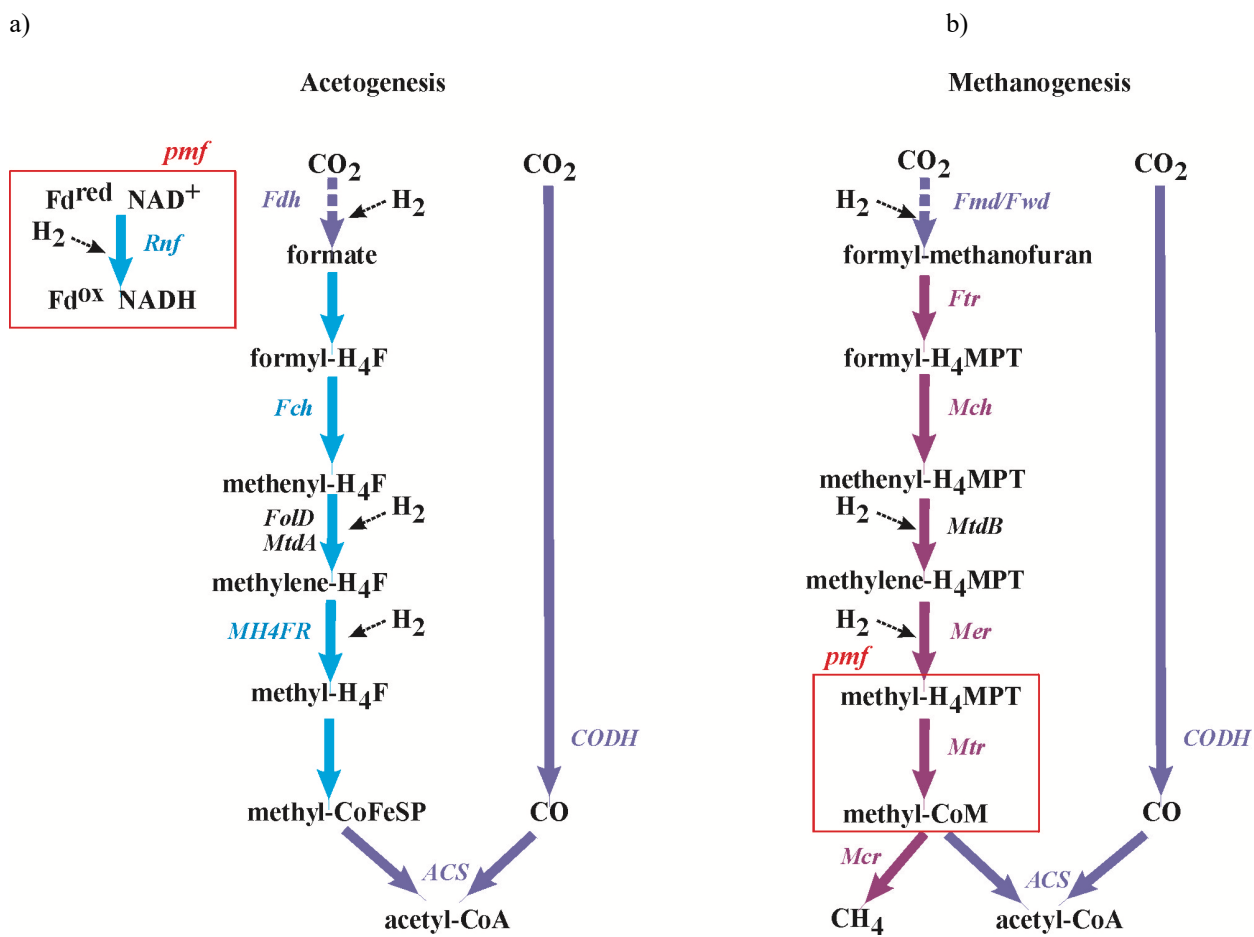

**Figure 4**. Adopted from (Nitschke and Russell 2013). Schematic representation of the WL pathways in acetogens (a) and methanogens (b). Steps restricted to Bacteria are marked in bright blue color, whereas those only found in Archaea are shown in violet. Dark blue stands for reactions and enzymes observed in both prokaryotic domains. Abbreviations are as follows: Fdh, formate dehydrogenase; Fmd/Fwd, Formylmethanofuran dehydrogenase; Mtr, coenzyme M methyltransferase; Mcr, methyl-coenzyme M reductase; Mer, methyl- ene-H4MPT reductase; Mch, methenyl-H4MPT

cyclohydrolase; Ftr, formylmethanofuran:H4MPT formyltransferase; Fch, methenyl- H4F cyclohydrolase; CODH, carbon monoxide dehydrogenase; ACS, acetyl coenzyme A synthase

In the Methyl branch $CO_2$ is reduced through a series of sequential hydrogenations reactions to methyl intermediate (Karekar and Stefanini 2022). In the carbonyl branch, $CO_2$ is first reduced to carbon monoxide (CO) intermediate by means of the enzyme carbon monoxide dehydrogenase (CODH). The CO then is conjugated to the Methyl intermediate by means of acetyl-CoA synthase (ACS) to produce acetyl coenzyme A (acetyl-CoA). This acetyl-CoA is subsequently converted to acetate during catabolism or utilized in the synthesis of cell carbon during anabolism (Rosenberg et al. 2013).

If the WL pathway is compared between methanogenic archaea and acetogens it is found that; in archaea the route that leads to methanogenesis is an extra extension and involves additional enzymes i.e. coenzyme M methyltransferase *Mtr*; methyl-coenzyme M reductase *Mcr*; collectively known as the *Mtr-Mcr* "module," **Fig. 4**; (Scheller and Rother 2022). This distinction apparently present credentials potentially for microbial metabolism manipulation i.e., we can simplify or deconstruct the pathway to acetogenesis. So, all that is needed is delete, suppress or by pass the route that leads to methane production. This can be achieved through spontaneous mutation or tools for genome editing.

Since acetyl CoA is believed to be an ancient metabolism and in the evolutionary progression, methanogens came relatively late than acetogens because of the energy conservation benefit (Martin 2020), experiments to revert methanogenic archaea to acetogens seems plausible. Literature shows that the function of the WL pathway changes with cultivation condition of a cell (Rosenberg et al. 2013). Thus, methanogenic metabolism can be rewired by just diverging all substrate carbon from methanogenesis to flow through acetyl-CoA, and convert methanogens into acetogens. This can be achieved either by growing methanogens on a medium that lacks key substrates for methanogenesis or grow them with an inhibitor for methanogenesis. This has been experimentally tested with cytochrome-containing methanogens like *Methanosarcina acetivorans* that uses acetate or methyl to form methane. The experiment proved that growth on carbon monoxide and two inhibitors; 2-bromoethanesulfonate (BES) and 3-nitrooxypropanol (3- NOP), lead to acetate production (Scheller and Rother 2022). Despite this milestone, acetoclastic methanogens in the rumen are very few (Nagarajan, Lee, and Chang 2019) and their contribution to methane is insignificant.

Furthermore, research has demonstrated a notable reduction in the transcript abundance of genes responsible for enzymes involved in the hydrogenotrophic methanogenesis pathway within the rumen of sheep that exhibit low CH4 emissions, as compared to sheep displaying high CH4 emissions (Shi et al. 2014). It is thus reasonable to hypothesize that the down-regulation of specific genes associated with hydrogenotrophic methanogenesis could serve as a valuable strategy for mitigating CH4 emissions. The genes responsible for encoding the crucial *Mcr* enzyme could be targeted because they are present in all methanogens and catalyzes the final step in methanogenesis.

The exceptional precision of CRISPR/Cas-mediated gene editing technology stems from the single guide RNA (sgRNA) that directs Cas nucleases to the target site within the host genome. If these sgRNA are well designed off-target mutations are significantly reduced. Therefore, alternative to the spontaneous mutation described above, CRISPR/Cas can be employed to target the genes that leads to methanogenesis i.e., *Mtr*-encoding genes and *Mcr* genes and arrive at the same goal. In fact, recently discovered methanogens that are capable of acetogenesis do not contain the *Mcr* genes (Orsi et al. 2020)

The availability of the genome sequence of *M. ruminantium* (Leahy et al. 2010) and the recent identification of its operome functional properties (Bharathi, Senthil Kumar, and Chellapandi 2020), combined with the extensive understanding of methanogenic pathways' biochemistry (Deppenmeier 2002), makes this approach a possibility. It is intriguing to note that advancements in organism improvements through gene editing techniques, specifically site-directed nucleases (SDN) like SDN-1, which are commonly utilized for gene knock-out or knock-down purposes, are not automatically categorized as 'genetically modified' (GM) in certain countries and have a totally different ethical landscape. Mainly because the mutations originate from organisms' own DNA repair mechanisms rather than an added foreign DNA sequence. Consequently, these gene editing approaches are not subjected to regulatory oversight in countries such as the United States, Canada, Brazil, Argentina, Israel, Nigeria, Australia, and Japan, among others (Subedi et al. 2022)

Attempts to delete *Mcr* gene has been done and the deletions using native homology-dependent repair (HDR)-mediated CRISPR/Cas in genes responsible for encoding an *Mcr* enzyme and a heterodisulfide reductase (*hdrED*) delivered only five transformants in *M. acetivorans* (Nayak and Metcalf 2017). *Mcr* and *hdrED* are essential genes and their deletion is expected to be lethal. Conversely, when deletions were made in non-essential genes responsible for encoding monomethylamina-specific methyltransferases (*mtmCB1* and *mtmCB2*), which facilitate methylamine utilization in the methylotrophic pathway, thousands of transformants were generated (Nayak and Metcalf 2017). These findings suggest that targeting methyltransferase genes for knockout may be a more favorable approach when aiming to decrease CH4 production in the rumen, particularly in the context of ruminant methanogens that utilize the methylotrophic pathway.

To study essential genes however, CRISPR interference system (CRISPRi) can be used to achieve reductions in gene expression. This method offers advantages when targeting essential genes like *Mcr*, as it does not involve DNA strand breaks and effects are reversible (Summary 2014). For instance, the expression of a target gene can be suppressed by utilizing an inactive dead Cas9 (dCas9) lacking endonucleolytic activity. Along with a single-guide RNA (sgRNA), dCas9 binds to the target DNA sequence within a protein-coding region, thereby obstructing RNA polymerase and transcript elongation (Qi et al. 2013; Mandegar et al. 2016). In *M. acetivorans*, the CRISPRi system has been effectively incorporated into the genome or transformed into a replicating plasmid, leading to a significant decrease of up to 90% in the transcript abundance of the target gene (Ahmed E. Dhamad 2020).

Similar to the dCas9 approach an endogenous Cas protein that specifically targets RNAs instead of double-stranded DNA cleavage can be employed. This approach is exemplified in the class 1 type III-B CRISPR/Cas system that can be found in few archaea (Hale et al. 2014). This approach has demonstrated successful application in *Sulfolobus solfataricus*, where specific RNAs were effectively down-regulated by harnessing an endogenous endonuclease in conjunction with an introduced guide small CRISPR RNA (crRNA). Theoretically, this strategy could also be extended to other archaea, providing a potential avenue for targeted RNA modulation in a broader range of organisms (Zebec et al. 2014; Zink et al. 2021). Nevertheless, there is considerable variation in CRISPR/Cas subtypes among methanogens, even among strains within the same species. As a result, a genome-editing technique that relies on the native CRISPR/Cas machinery for one strain may not be effective in other closely related strains. It will be of great interest to see how these technologies perform in *M. ruminantium*, the dominant methane producer in ruminant animals.

The potential of using engineered organisms to reduce methane emissions in cattle is promising, as these strains may be more effective in reducing methane compared to traditional breeding for low emitters. However, successful integration of these engineered organisms into the rumen microbiome of all cattle may present a challenge. Assuming the engineered strain can reduce methane by 50% in treated cattle, it is important to consider the feasibility of spreading the modified organisms throughout the entire cattle population. Hypothetical calculations can be made to determine the time frame necessary to achieve this goal, taking into account factors such as animal turnover rate, reproduction rates, and the potential for horizontal transfer of modified organisms.

**Conclusion**

Methane is a potent greenhouse gas. A large part of emission is microbial in origin where ruminant enteric fermentation is the biggest contributor. Interventions that target methane emission are effective in reducing greenhouse effects in the short term and livestock is a ripe sector for attack. Some of the proposed strategies such as eliminating or reducing livestock production are akin to throwing the baby with the bathwater and don't address the root source of emission. In this paper we put emphasis on biotechnological solutions targeting rumen microbes that are likely to be more effective, efficient and timely. Moreover, they are likely to be scaled to other biogenic sources such as wetlands and rice paddies. The problem with anaerobic fermentation is that it generates hydrogen which when not given an alternative is consumed in methane production. We postulate that the WL pathway especially channeling $H_2$ to acetogenesis can be a remedy to rumen methane production and provide a pervasive avenue for addressing global methane emission beyond the rumen ecosystem.

**Funding Sources**

We appreciate the support of Regional Scholarship and Innovation Fund (RSIF) through the Partnership for Skills in Applied Sciences, Engineering and Technology (PASET) program for funding this project. The funders were not involved in preparation of this article.

**Author contributions**

**Rehema Mrutu:** Conceptualization, Writing- Original draft preparation, **Abdussamad Muhammad Abdussamad**: Supervision, Reviewing and Editing. **Kabir Mustapha Umar**: Reviewing and Editing. **Morris Agaba**: Reviewing and Editing **Adnan Abdulhamid**: Reviewing and Editing.

**Competing interests**

The authors have no competing interests to declare that are relevant to the content of this article.

26Gagen EJ, Denman SE, Padmanabha J, Zadbuke S, Jassim R Al, Morrison M, McSweeney CS (2010) Functional gene analysis suggests different acetogen populations in the bovine rumen and tammar wallaby forestomach. Applied and Environmental Microbiology 76:7785–7795. https://doi.org/10.1128/AEM.01679-10

Gagen EJ, Wang J, Padmanabha J, Liu J, De Carvalho I PC, Liu J, Webb RI, Al Jassim R, Morrison M, Denman S E, McSweeney CS (2014) Investigation of a new acetogen isolated from an enrichment of the tammar wallaby forestomach. BMC Microbiology 14: 1–14. https://doi.org/10.1186/s12866-014-0314-3

Georges M, Charlier C, Hayes B (2019) Harnessing genomic information for livestock improvement. In *Nature* Reviews Genetics 20:135–156. https://doi.org/10.1038/s41576-018-0082-2

Gilbert RA, Kelly WJ, Altermann E, Leahy SC, Minchin C, Ouwerkerk D, Klieve AV (2017) Toward understanding phage: Host interactions in the rumen; complete genome sequences of lytic phages infecting rumen bacteria. Frontiers in Microbiology 8: 1–17. https://doi.org/10.3389/fmicb.2017.02340

Golder HM, Thomson JM, Denman SE, McSweeney CS, Lean IJ (2018) Genetic markers are associated with the ruminal microbiome and metabolome in grain and sugar challenged dairy heifers. Frontiers in Genetics 9: 1–10. https://doi.org/10.3389/fgene.2018.00062

González-Recio O, López-Paredes J, Ouatahar L, Charfeddine N, Ugarte E, Alenda R, Jiménez-Montero JA (2020) Mitigation of greenhouse gases in dairy cattle via genetic selection: 2. Incorporating methane emissions into the breeding goal. Journal of Dairy Science 103: 7210–7221. https://doi.org/10.3168/jds.2019-17598

Granja-Salcedo YT, Fernandes RMI, De Araujo RC, Kishi LT, Berchielli TT, De Resende FD, Berndt A, Siqueira GR (2019) Long-term encapsulated nitrate supplementation modulates rumen microbial diversity and rumen fermentation to reduce methane emission in grazing steers. Frontiers in Microbiology 10: 1–12. https://doi.org/10.3389/fmicb.2019.00614

Hale CR, Alexis C, Hong L, Rebecca MT, Michael PT (2014) Target RNA Capture and Cleavage by the Cmr Type III-B CRISPR–Cas Effector Complex. Genes and Development 28 (21): 2432–43. https://doi.org/10.1101/gad.250712.114.

Hayek MN, Garrett RD (2018) Nationwide shift to grass-fed beef requires larger cattle population. *Environmental* Research Letters 13: 1–9. https://doi.org/10.1088/1748-9326/aad401

Humpenöder F, Bodirsky BL, Weindl I, Lotze-campen H, Linder T, Popp A (2022) Projected environmental benefits of replacing beef with microbial protein. Nature 10: 1–21. https://doi.org/10.1038/s41586-022-04629-

Ibrahim NA, Alimon AR, Yaakub H, Samsudin AA, Candyrine SCL, Wan Mohamed W N, Md Noh A, Fuat MA, Mookiah S (2021) Effects of vegetable oil supplementation on rumen fermentation and microbial population in ruminant: a review. Tropical Animal Health and Production 53:1–11. https://doi.org/10.1007/s11250-021-02863-4
26

Wallace RJ, Rooke JA, McKain N, Duthie CA, Hyslop JJ, Ross DW, Waterhouse A, Watson M, Roehe R (2015) The rumen microbial metagenome associated with high methane production in cattle. BMC Genomics 16: 1–14. https://doi.org/10.1186/s12864-015-2032-0

Weidenbach K, Nickel L, Neve H, Alkhnbashi OS, Künzel S, Kupczok A, Bauersachs T, Cassidy L, Tholey A, Backofen R, Schmitz RA (2017) Methanosarcina Spherical Virus, a Novel Archaeal Lytic Virus Targeting Methanosarcina Strains. Journal of Virology 91: 1–17. https://doi.org/10.1128/jvi.00955-17

Wesemael DVan, Vandaele L, Ampe B, Cattrysse H, Duval S, Kindermann M, Fievez V, Campeneere SDe, Peiren N (2019) Reducing enteric methane emissions from dairy cattle : Two ways to supplement 3-nitrooxypropanol. Journal of Dairy Science 102: 1780–1787. https://doi.org/10.3168/jds.2018-14534

Wilkinson T, Korir D, Ogugo M, Stewart RD, Watson M, Paxton E, Goopy J, Robert C (2020) 1200 high-quality metagenome-assembled genomes from the rumen of African cattle and their relevance in the context of sub-optimal feeding. Genome Biology 21: 1–25. https://doi.org/10.1186/s13059-020-02144-7

Willett W, Rockström J, Loken B, Springmann M, Lang T, Vermeulen S, Garnett T, Tilman D, DeClerck F, Wood, A, Jonell M, Clark M, Gordon LJ, Fanzo J, Hawkes C, Zurayk R, Rivera JA, De Vries W, Majele Sibanda L, Murray CJL (2019) Food in the Anthropocene: the EAT–Lancet Commission on healthy diets from sustainable food systems. The Lancet 393:447-492. https://doi.org/10.1016/S0140-6736(18)31788-4

Williams SRO, Hannah MC, Eckard RJ, Wales WJ, Moate PJ (2020) Supplementing the diet of dairy cows with fat or tannin reduces methane yield, and additively when fed in combination. Animal 14:64–72. https://doi.org/10.1017/S1751731120001032

Wilson, A. S., Koller, K. R., Ramaboli, M. C., Nesengani, L. T., Ocvirk, S., Chen, C., Flanagan, C. A., Sapp, F. R., Merritt, Z. T., Bhatti, F., Thomas, T. K., & O'Keefe, S. J. D. (2020). Diet and the Human Gut Microbiome: An International Review. In *Digestive Diseases and Sciences* (Vol. 65, Issue 3, pp. 723–740). Springer. https://doi.org/10.1007/s10620-020-06112-w

Xia Y, Kong YH, Seviour R, Forster RJ, Kisidayova S, Mcallister TA (2014) Fluorescence in situ hybridization probing of protozoal Entodinium spp. and their methanogenic colonizers in the rumen of cattle fed alfalfa hay or triticale straw. Journal of Applied Microbiology 116: 14–22. https://doi.org/10.1111/jam.12356

Xue MY, Sun HZ, Wu XH, Liu JX, Guan LL (2020) Multi-omics reveals that the rumen microbiome and its metabolome together with the host metabolome contribute to individualized dairy cow performance. Microbiome 8: 1–19. https://doi.org/10.1186/s40168-020-00819-8

Yáñez-Ruiz DR, Abecia L, Newbold CJ (2015) Manipulating rumen microbiome and fermentation through interventions during early life. Frontiers in Microbiology 6: 1–12. https://doi.org/10.3389/fmicb.2015.01133

Yang C (2018) Acetogen communities in the gut of herbivores and their potential role in syngas fermentation. Fermentation 4: 1–17. https://doi.org/10.3390/fermentation4020040